# CONTINUOUS – TIME RANDOM WALK WITH CORRELATED WAITING TIMES

Aleksei V. Chechkin,[1,2] Michael Hofmann[3], and Igor M. Sokolov[3]

[1]School of Chemistry, Tel Aviv University, Ramat Aviv, 69978 Tel Aviv, Israel

[2]Akhiezer Institute for Theoretical Physics NSC KIPT, Akademicheskaya st.1, 61108 Kharkov, Ukraine

[3]Institute of Physics, Humboldt University, Newtonstr.15, 12489 Berlin, Germany

**Abstract.**

Based on the Langevin description of the Continuous Time Random Walk (CTRW), we consider a generalization of CTRW in which the waiting times between the subsequent jumps are correlated. We discuss the cases of exponential and slowly decaying persistent power – law correlations between the waiting times as two generic examples and obtain the corresponding mean squared displacements as functions of time. In the case of exponential – type correlations the (sub)diffusion at short times is slower than in the absence of correlations. At long times the behavior of the mean squared displacement is the same as in uncorrelated CTRW. For power – law correlations we find subdiffusion characterized by the same exponent at all times, which appears to be smaller than the one in uncorrelated CTRW. Interestingly, in the limiting case of an extremely long power – law correlations, the (sub)diffusion exponent does not tend to zero, but is bounded from below by the subdiffusion exponent corresponding to a short – time behavior in the case of exponential correlations.

# I. INTRODUCTION

The Continuous Time Random Walk (CTRW) model was originally introduced by Montroll and Weiss in their seminal paper of 1965 [1]. Since that time the CTRW was proved to be a useful tool for the description of systems out of equilibrium, especially of anomalous diffusion phenomena characterized by non – linear time dependence of mean squared displacement (MSD), see, e.g., the reviews [2], [3], [4] and references therein. Within the simplest CTRW picture, the "jump model", diffusion of a particle is considered as a sequence of independent random jumps occurring instantaneously; the waiting times between the successive jumps are independent random variables. Thus, the motion of a particle is completely determined by the two probability density functions (PDFs), namely, jump length PDF and waiting – time (or pausing – time) PDF. Other CTRW schemes (velocity and two – state models) account for the constant velocity between the points of halt [5], [6].

In the present paper we concentrate on the particular case of the jump CTRW model where the distribution of waiting times $\Delta t_n$, $n = 1, 2, ...$ between the jumps (given by its probability density $\psi(\Delta t)$) possesses long power – law asymptotics, $\psi(\Delta t) \propto \Delta t^{1+\alpha}$, $0 < \alpha < 1$, such that the mean waiting time diverges, whereas the variance $a^2$ of the jump length is finite. This case goes back to the paper by Scher and Montroll on amorphous semiconductors [7], and leads to subdiffusion of a particle, that is, the MSD growing sublinearly with time. Besides the probabilistic approach following the lines of Montroll and Weiss' work [1], CTRW can be described using Generalized Master Equations and within the Langevin approach. It was shown [8] that the Generalized Master Equation for the jump CTRW theory is reduced to a time – fractional Fokker – Planck equation in the long time limit. Since then the theory of time – fractional kinetic equations was intensively developed, see the reviews [9], [10], [11]. On the other hand, Fogedby [12] introduced a Langevin approach to time – fractional dynamics and demonstrated how time – fractional diffusion equation arises from the corresponding coupled Langevin equations. Recently, this approach was further developed in a series of papers by Friedrich et al [13], [14], [15], [16].

The purpose of the present work is to generalize the CTRW approach by including correlations between the waiting times. This is in line with the recent studies of the CTRW with correlated jumps [17]. Apart from purely theoretical interest, such model is motivated by possible existence of such correlations in biological movements [18] and in financial applications of CTRW [19]. As a tool to consider the effect of correlations between the waiting times we use the Langevin approach. The structure of the paper is as follows. The Langevin approach to uncorrelated CTRW is recalled briefly in Section II. In Section III we propose its generalization by including memory effects into the Langevin description and get a general expression for the MSD. In subsequent Sections IV and V we consider the particular examples of exponential and power law correlations between waiting times and derive the MSD for both cases. In Section VI we present results of numerical simulations corroborating the ones of analytical consideration. For the reader's convenience, some details of the derivation are shifted to the Appendices A – C.

## II.   LANGEVIN APPROACH TO SUBDIFFUSIVE CTRW MODEL

In this Section we recall briefly the subdiffusive jump CTRW model [20] and the corresponding Langevin approach [12], [13], [14]. The key element of the CTRW is the subordination of random processes [21]. For example, for the PDF to find a particle (with continuous distribution of step lengths) at point $x$ at time $t$ one has

$$f(x,t) = \sum_n f_1(x,n) h_n(t) \; , \tag{1}$$

where $f_1(x,n)$ is a PDF to find a particle at point $x$ after $n$ steps, and $h_n(t)$ is the probability to make exactly $n$ steps up to time $t$. The function $h_n(t)$ is connected to the waiting time PDF $\psi(\Delta t)$ via their Laplace transforms $\tilde{h}_n(u) = (1-\tilde{\psi}(u))\tilde{\psi}^n(u)/u$, see, e.g., [20]. Here, the random number of steps plays the role of internal, operational time governing the system's evolution. In the classical Scher – Montroll picture $h_n(t)$ corresponds to a random process in which $n$ grows sublinearly in $t$, thus the operational time is always in delay compared with the physical one, and the overall process $x(n(t))$ is subdiffusive. From Eq.(1) it follows e.g. that the mean squared displacement $\langle x^2(t) \rangle$ is essentially a function of the mean number of steps,

$$\langle x^2(t) \rangle = a^2 \langle n(t) \rangle = a^2 N(t) \quad , \tag{2}$$

where the mean number of steps performed up to time $t$, $N(t)$, is defined as $N(t) = \sum_{n=0}^{\infty} n h_n(t)$.

The CTRW process is semi-Markovian in the sense that at each new step a displacement $\Delta x$ and the corresponding waiting time $\Delta t > 0$ are chosen at random from the corresponding probability distributions and are independent on the prehistory of the process. The overall displacement after $n$-th step and the (physical) time of $(n+1)$-st step are given by the sums of the corresponding increments, thus the physical time $t$ of jump and the displacement $x(t)$ are Markov chains in the internal variable $n$.

The Fogedby's approach [12] expresses this property in a continuum approximation via the coupled Langevin equations with the two stationary noise sources,

$$\frac{dx(s)}{ds} = \eta(s) \quad , \tag{3}$$

$$\frac{dt(s)}{ds} = \tau(s) \quad , \tag{4}$$

where the random walk $x(t)$ is parametrized in terms of a continuous variable $s$, which actually has a meaning of operational time: the process $s(t)$ is a continuum analogue of the number of steps $n(t)$.

Now, we specify the random processes $x(s)$ and $t(s)$ by explicitly defining the properties of the noise sources in Eqs.(3) and (4). Namely, the stationary random process $\eta(s)$ is a white Gaussian noise, $\langle \eta(s) \rangle = 0$, $\langle \eta(s)\eta(s') \rangle = 2\delta(s-s')$, thus the process $x(s) = \int_0^s ds' \eta(s')$ is a Wiener process with the PDF

$$f_1(x,s) = \frac{1}{\sqrt{4\pi s}} \exp\left(-\frac{x^2}{4s}\right) \tag{5}$$

(we take $x(0)=0$). In its turn, the stationary random process $\tau(s)$ is a white alpha – stable Lévy noise, which takes positive values only. That is, the process

$$t(s) = \int_0^s ds' \tau(s') \tag{6}$$

is an $\alpha$ - stable totally skewed Lévy motion with the Lévy index $\alpha$, $0 < \alpha < 1$. The characteristic function $\hat{L}_\alpha(k,s)$, that is the Fourier transform of the PDF $L_\alpha(t,s)$, has the form

$$\hat{L}_\alpha(k,s) = \int_0^\infty dt\, e^{ikt} L_\alpha(t,s) = \exp\left\{-s\,|k|^\alpha \exp\left(-\frac{i\pi\alpha}{2}\operatorname{sgn} k\right)\right\}. \tag{7}$$

Here we use Feller – Takayasu canonical form for strictly stable distributions, see [21], [22], [23]. The Laplace transform $\tilde{L}_\alpha(\lambda,s)$ of the PDF $L_\alpha(t,s)$ reads [21], [24],

$$\tilde{L}_\alpha(\lambda,s) = \int_0^\infty dt\, e^{-\lambda t} L_\alpha(t,s) = \exp\{-\lambda^\alpha s\}\,, \quad 0 < \alpha < 1. \tag{8}$$

For the sake of simplicity we consider both noises $\eta$ and $\tau$ to have unit intensities.

The PDF $f(x,t)$ of the process $x(t)$ is then given by

$$f(x,t) = \int_0^\infty ds\, f_1(x,s) h(s,t) \;, \tag{9}$$

being the continuous analogue of Eq.(1). Here $h(s,t)$ is the PDF of random variable $s$ at time $t$. The random function $s(t)$ is inverse to $t(s)$ defined by Eq.(6). To determine $h(s,t)$, we note that the random function $t(s)$ is monotonical,

$$s_2 > s_1 \Rightarrow t(s_2) > t(s_1) \;, \tag{10}$$

so that

$$\Theta(s - s(t)) = 1 - \Theta(t - t(s)) \;, \tag{11}$$

where $\Theta(x)$ is the Heaviside step function, $\Theta(x) = 1$ for $x > 0$, $\Theta(x) = 0$ for $x < 0$, $\Theta(x=0) = 1/2$. Statistical averaging of Eq.(11) gives

$$\langle \Theta(s - s(t)) \rangle = 1 - \langle \Theta(t - t(s)) \rangle \;. \tag{12}$$

The PDF $h(s,t)$ is therefore written as

$$h(s,t) = \langle \delta(s - s(t)) \rangle = \frac{\partial}{\partial s} \langle \Theta(s - s(t)) \rangle = -\frac{\partial}{\partial s} \langle \Theta(t - t(s)) \rangle \;. \tag{13}$$

For the Laplace transform $\tilde{h}(s, \lambda)$ we get, with the use of Eq.(8),

$$\tilde{h}(s,\lambda) = -\frac{\partial}{\partial s}\left\langle \int_0^\infty dt\, e^{-\lambda t}\Theta(t-t(s))\right\rangle = -\frac{\partial}{\partial s}\frac{1}{\lambda}\left\langle \int_0^\infty dt\, e^{-\lambda t}\delta(t-t(s))\right\rangle =$$

$$= -\frac{\partial}{\partial s}\frac{1}{\lambda}\tilde{L}_\alpha(\lambda,s) = \lambda^{\alpha-1}e^{-s\lambda^\alpha} \quad . \tag{14}$$

The inverse Laplace transform of $\tilde{h}(s,\lambda)$ has been found in Ref.[25], see Eqs.(16) and (24) there,

$$h(s,t) = \frac{t}{\alpha s^{1+1/\alpha}} L_\alpha\left(\frac{t}{s^{1/\alpha}}\right) \quad , \tag{15}$$

where $L_\alpha(y)$ is a one sided Lévy stable PDF whose Laplace transform is $\tilde{L}_\alpha(u) = \exp(-u^\alpha)$. Equation (15) together with Eq.(5) defines the PDF $f(x,t)$ via Eq.(9).

In our paper we are interested in the MSD,

$$\langle x^2 \rangle(t) = \int_{-\infty}^{\infty} dx\, x^2 f(x,t) = 2\int_0^\infty ds\, s\, h(s,t) \quad . \tag{16}$$

With changing the variable of integration, $s \to y = t/s^{1/\alpha}$, Eq.(16) reads

$$\langle x^2 \rangle(t) = 2t^\alpha \int_0^\infty dy\, y^{-\alpha} L_\alpha(y) \quad . \tag{17}$$

Using the formula (A.15), we arrive at

$$\langle x^2 \rangle(t) = \frac{2}{\Gamma(1+\alpha)} t^{\alpha} , \tag{18}$$

which is a known result from the theory of subdiffusive CTRW [8], [26].

## III. CORRELATED PROBLEM

Let us now consider the CTRW where the waiting times between the subsequent steps are not independent. The simplest way to introduce such dependence is to assume the corresponding waiting times to be weighted sums of independent random variables:

$$\Delta t_n = \sum_{j=0}^{n} M_{n,j} \tau_j \tag{19}$$

where $M_{n,j} = M(n-j)$ is a memory function. Independent random variables $\tau_j$ are identically distributed with PDF $\psi(\tau)$; the uncorrelated case corresponds to $M_{n,j} = \delta_{nj}$. In what follows we take $\psi(\tau)$ to be a one – sided Lévy stable PDF $L_\alpha(\tau)$, as above. Two types of memory functions are considered: An exponential one, given by geometric series, $M(k) = (1-q)q^k$ with $0 < q < 1$ (normalized), which corresponds to $M(k) = \Delta^{-1} \exp(-k/\Delta)$ with $\Delta \approx (1-q)^{-1}$ for $q$ close to unity, and a power-law one $M(k) = k^{-\beta}$ with $0 < \beta < 1$ (not normalizable).

In the continuous model the corresponding memory can be introduced also into the Langevin equation for $t(s)$ (compare with Eq.(4)),

$$\frac{dt(s)}{ds} = \int_0^s ds' M(s-s') \tau(s') , \tag{20}$$

where $M$ is a non-negative continuous memory function, and $\tau(s)$ are the same noises as in Eq.(4). The random process $x(s)$ is still defined by Eq.(3).

The exponential memory function in the first case corresponds now to

$$M(s) = \frac{1}{\Delta}\exp\left(-\frac{s}{\Delta}\right) . \tag{21}$$

The power-law function corresponds to $M(s) \propto s^{-\beta}$. In particular, we use

$$M(s) = \frac{s^{-\beta}}{\Gamma(1-\beta)} , \quad 0 < \beta < 1 , \tag{22}$$

in order to allow for a continuous limiting transition to a non-correlated case at $\beta \to 1$.

Integrating Eq.(20) we get

$$t(s) = \int_0^s ds' \int_0^{s'} ds'' M(s'-s'')\tau(s'') = \int_0^s ds' \tau(s')\mu(s,s') , \tag{23}$$

where

$$\mu(s,s') = \int_{s'}^s ds'' M(s''-s') . \tag{24}$$

The characteristic function $\hat{p}(k,s)$ of the process $t(s)$ takes the form, see Appendix B,

$$\hat{p}(k,s) = \int_{-\infty}^{\infty} dt\, e^{ikt} p(t,s) = \exp\left\{-|k|^\alpha \phi(s)\exp\left(-\frac{i\pi\alpha}{2}\operatorname{sgn} k\right)\right\}, \quad 0 < \alpha < 1 , \tag{25}$$

where

$$\phi(s) = \int_0^s ds' \left( \int_{s'}^s ds'' M(s''-s') \right)^\alpha . \qquad (26)$$

One can see that the variable $s$ in Eq.(7) is substituted by $\phi(s)$ in Eq.(25). Correspondingly, for the Laplace transform $\tilde{p}(\lambda,s)$ of the PDF $p(t,s)$ we get

$$\tilde{p}(\lambda,s) = \int_0^\infty dt\, e^{-\lambda t} p(t,s) = \exp\{-\lambda^\alpha \phi(s)\} . \qquad (27)$$

Now, we simply go along the lines of Section II. For the function $\tilde{h}(s,\lambda)$ we get

$$\tilde{h}(s,\lambda) = -\frac{\partial}{\partial s} \frac{1}{\lambda} \tilde{p}(\lambda,s) = \lambda^{\alpha-1} \phi'(s) \tilde{p}(\lambda,s) \qquad (28)$$

(compare with Eq.(14)), and after the inverse Laplace transform $\lambda \to t$ we obtain

$$h(s,t) = \frac{\phi'(s)}{\alpha} \frac{t}{(\phi(s))^{1+1/\alpha}} L_\alpha\left( \frac{t}{(\phi(s))^{1/\alpha}} \right). \qquad (29)$$

The MSD is obtained by inserting Eq.(29) into Eq.(16),

$$\langle x^2 \rangle(t) = \frac{2}{\alpha} \int_0^\infty ds\, s\, \phi'(s) \frac{t}{[\phi(s)]^{1+1/\alpha}} L_\alpha\left[ \frac{t}{(\phi(s))^{1/\alpha}} \right] . \qquad (30)$$

Changing variable $s \to y = t/(\phi(s))^{1/\alpha}$, we get

$$\langle x^2 \rangle(t) = 2\int_0^\infty dy\, \phi^{-1}\left(\frac{t^\alpha}{y^\alpha}\right) L_\alpha(y) \;, \tag{31}$$

where $\phi^{-1}(\xi)$ is an inverse function of $\phi(s)$, that is $\phi^{-1}(\phi(s)) = s$. In the uncorrelated limit $\phi(s) = s$, thus $\phi^{-1}(\xi) = \xi$, and Eq.(31) is equivalent to Eq.(17).

## IV.  EXPONENTIAL CORRELATIONS

We choose $M(s)$ as given by Eq.(21). Then, using Eq.(26) we get

$$\phi(s) = \int_0^s ds' \left[1 - e^{-s'/\Delta}\right]^\alpha \;. \tag{32}$$

Now we consider the MSD in the limits of short and long times, respectively.

*Short times.* At small values of $s$, $s < \Delta$, we approximate $\phi(s)$ by

$$\phi(s) \approx \int_0^s ds' \left(\frac{s'}{\Delta}\right)^\alpha = \frac{s^{1+\alpha}}{(1+\alpha)\Delta^\alpha} \;. \tag{33}$$

Thus,

$$\phi^{-1}(\xi) = \left[(1+\alpha)\Delta^{\alpha}\right]^{1/(1+\alpha)} \xi^{1/(1+\alpha)} \ . \tag{34}$$

Equation (31) then takes the form

$$\left\langle x^2 \right\rangle(t) = t^{\frac{\alpha}{1+\alpha}} 2(1+\alpha)^{1/(1+\alpha)} \Delta^{\alpha/(1+\alpha)} \int_0^{\infty} dy\, y^{-\frac{\alpha}{1+\alpha}} L_{\alpha}(y) \ . \tag{35}$$

With the use of Eq.(A.14) we get ultimately

$$\left\langle x^2 \right\rangle(t) = K_{\alpha} t^{\frac{\alpha}{1+\alpha}} \quad , t < \Delta \ , \tag{36}$$

where

$$K_{\alpha} = 2\Delta^{\alpha/(1+\alpha)} (1+\alpha)^{1/(1+\alpha)} \frac{\Gamma\left(\dfrac{1}{1+\alpha}\right)}{\alpha \Gamma\left(\dfrac{\alpha}{1+\alpha}\right)} \ . \tag{37}$$

We note that this result can be also obtained by the use of the Laplace method, see Appendix C.

***Long times.*** At long times, $s > \Delta$, Eq.(32) gives $\phi(s) \approx s$. Therefore, $\phi^{-1}(\xi) = \xi$, so that Eq.(31) gives

$$\left\langle x^2 \right\rangle(t) = 2t^{\alpha} \int_0^{\infty} dy\, y^{-\alpha} L_{\alpha}(y) = \frac{2t^{\alpha}}{\Gamma(1+\alpha)} \ , \tag{38}$$

see Eq.(A15). This result coincides with that obtained in an uncorrelated CTRW, see Eq.(18).

## V. POWER – LAW CORRELATIONS

Now we consider power-law memory kernel as given by Eq.(22). Equation (26) then leads to

$$\phi(s) = \frac{1}{[\Gamma(2-\beta)]^\alpha} \frac{s^\nu}{\nu} \ , \tag{39}$$

where

$$\alpha(1-\beta) + 1 = \nu \ , \tag{40}$$

and therefore

$$\phi^{-1}(\xi) = \xi^{1/\nu} [\Gamma(2-\beta)]^{\alpha/\nu} \nu^{1/\nu} \ . \tag{41}$$

Inserting Eq.(41) into Eq.(31) gives

$$\left\langle x^2(t) \right\rangle = 2t^{\alpha/\nu} [\Gamma(2-\beta)]^{\alpha/\nu} \nu^{1/\nu} \int_0^\infty dy \, y^{-\alpha/\nu} L_\alpha(y) \ . \tag{42}$$

With the use of Eq.(A14) we get

$$\left\langle x^2 \right\rangle(t) = K_{\alpha,\beta} t^{\alpha/\nu} \ , \tag{43}$$

where

$$K_{\alpha,\beta} = 2[\Gamma(2-\beta)]^{\alpha/\nu} \frac{\nu^{1/\nu}\Gamma\left(\frac{1}{\nu}\right)}{\alpha\Gamma\left(\frac{\alpha}{\nu}\right)} \quad . \tag{44}$$

In the case of normal diffusion $\alpha = \beta = 1$ Eq.(44) gives $\langle x^2(t)\rangle = 2t$, as it should. In case of $\beta = 1, \alpha \neq 1$ Eqs.(43), (44) reproduce the result of Eq.(18). The case $\beta \neq 1$ leads to slower subdiffusion than in the uncorrelated case, with the exponent $\alpha/\nu$ being monotonic in $\beta$. In the limit of very strong correlations, $\beta = 0$, one obtains

$$\langle x^2\rangle(t) \sim t^{\alpha/(1+\alpha)} \quad , \tag{45}$$

i.e., the exponent of anomalous diffusion $\alpha/\nu$ is bounded from below and does not tend to zero even for $\beta \to 0$. Interestingly, this lower bound corresponds to the short time behavior in the case of exponential correlations, Eq.(36). In the limiting case $\alpha \to 1$ leading to a normal diffusion law in the absence of correlations, the correlations between waiting times lead to decrease of the time exponent from 1 for $\beta = 1$ to 1/2 for very strong correlations, $\beta = 0$.

## VI. NUMERICAL SIMULATIONS

The asymptotic results of our continuous approach are corroborated by numerical simulations of a discrete model. Since the mean squared displacement in the CTRW follows exactly the mean number of steps $N(t)$, see Eq.(2), we only present the results for this last quantity.

We thus generate the waiting times according to Eq.(19) with the corresponding memory functions $M(k)$. The results presented below correspond to $\alpha = 1/2$. The independent random variables $\tau_i$ following the one-sided Lévy – Smirnov distribution

$$\psi(\tau) = (2\pi)^{-1/2} \tau^{-3/2} \exp(-1/2\tau) \tag{46}$$

are generated using Janicky-Weron algorithm [27]. We note that Laplace transform of Eq.(46) reads as $\tilde{\psi}(u) = \exp\left(-(2u)^{1/2}\right)$, which differs by a factor $\sqrt{2}$ in the exponent from the form $\tilde{L}_\alpha(u) = \exp(-u^\alpha)$ adopted in our analytical calculations, see Sec.II. The factor $\sqrt{2}$ will be important for our plot in Fig.3. Equation (46) implies that the time is measured in units of the scale factor of the distribution which equals to one. From the sequences of jump times the numbers of jumps performed up to fixed output times $t_{out}$ are obtained and averaged over $10^6$ independent realizations of the process.

The behavior of $N(t)$ in the exponentially correlated process with $q = 0.9$ and for power-law correlations with three different values of $\beta = 0.5, 0.25$ and $0.1$ are presented in Fig.1. One readily infers that at long times the $N(t)$-dependence follows power laws $N(t) \propto t^\gamma$. The values of $\gamma$ are obtained as least square fits to the curves on double-logarithmic scales over the four last decades in time and equal to $\gamma = 0.50$ for the exponential correlations and $\gamma = 0.39, 0.36$ and $0.34$ for the given values of $\beta$, respectively. These values, within the uncertainty limits, coincide with theoretical predictions $\gamma = \alpha/[\alpha(1-\beta)+1]$ being $2/5 = 0.4$, $4/11 \approx 0.364$ and $10/29 \approx 0.349$.

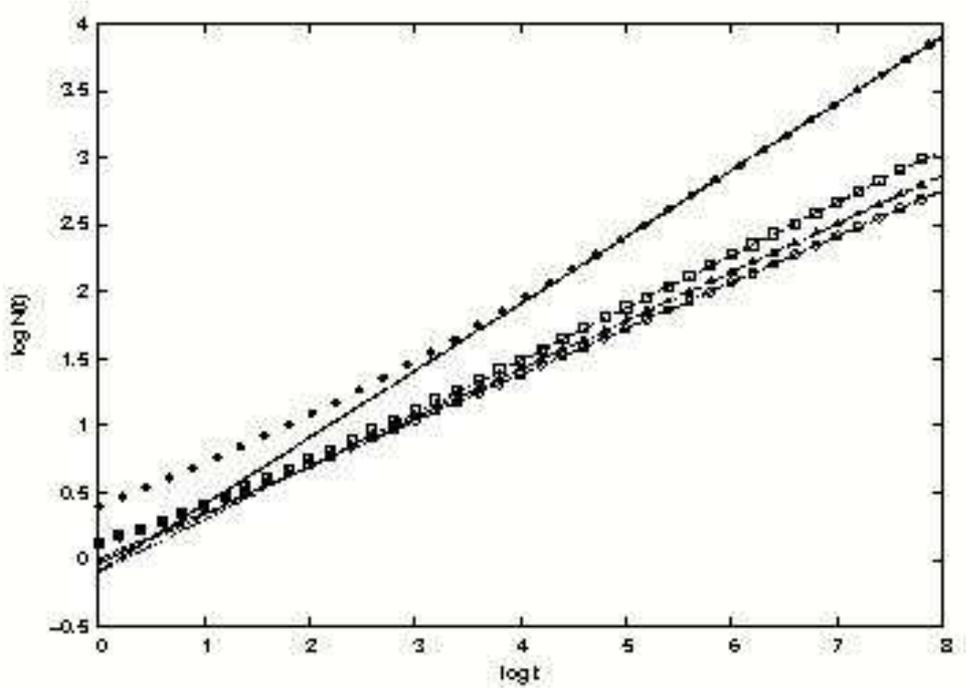

**Fig.1. Number of steps as a function of time on a log – log scale for exponential correlations with $q = 0.9$ (black diamonds) and for power – law correlations with $\beta = 0.5$ (empty rectangles), 0.25 (black triangles) and 0.1 (empty circles). The corresponding fitting lines have the slopes close to the analytical ones, see text for details.**

In Fig.2 the behavior of $N(t)$ is plotted versus $t$ on a log – log scale for $q = 0.999$. The short – and long time asymptotical behavior is clearly seen. Note that taking $q$ very close to unity is necessary since the short – time asymptotics persists only until $t \approx \Delta \approx (1-q)^{-1}$, see Eq.(36). The dotted line represents the least square fit to the data over the first three decades ($1 \leq t \leq 10^3$) in time and has a slope 0.33 (theoretical value 1/3). The full line represents the least square fit to the data over the last three decades ($10^8 \leq t \leq 10^{11}$) and has a slope 0.49 (theoretical value 1/2).

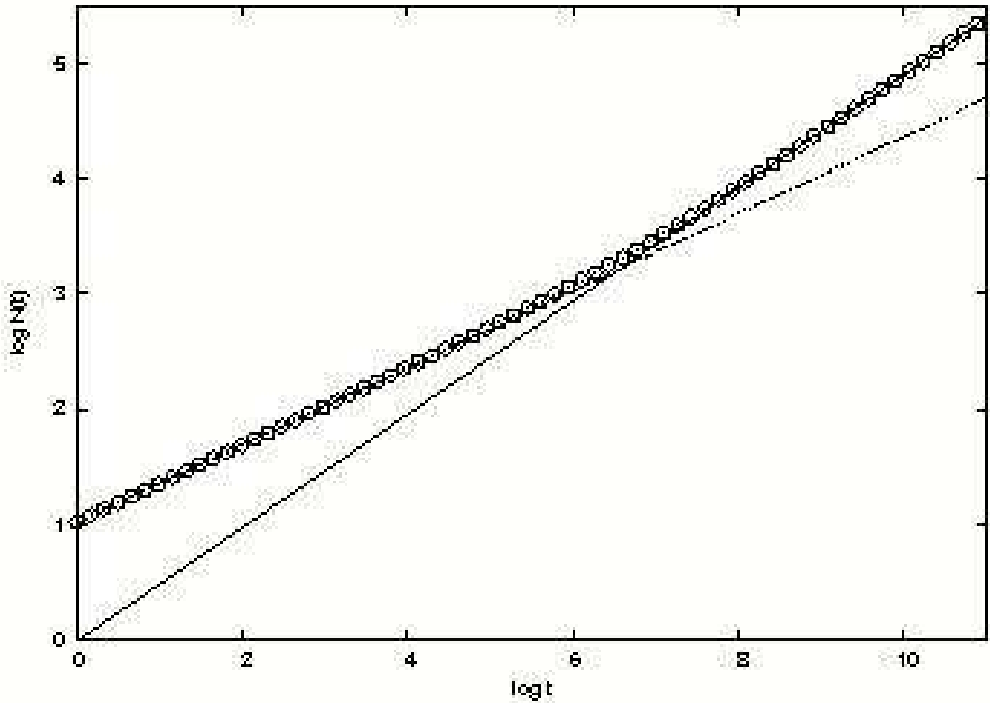

**Fig.2. Number of steps as a function of time on a log – log scale for exponential correlations with $q = 0.999$. The dotted line represents the power – law asymptotics at small times whereas the full line represents the asymptotics at large times. See text for details.**

Fig. 3 illustrates the behavior of $N(t)$ for exponential correlations with different values of $q = 0.5, 0.9, 0.99$ and $0.999$. Plotted is $A(t) = N(t)/t^\alpha$ as a function of time on a logarithmic scale. The fact that $A(t)$ converges to the same value (showing only fluctuations of the order of a few per mille) for all four different values of $q$ is clearly seen (note that the whole ordinate axis corresponds to the change of 3%). The asymptotic value of $A(t)$ (under the time units adopted) is equal to $1/\left(\sqrt{2}\Gamma(3/2)\right) = \sqrt{2/\pi} = 0.798$, which is shown in Fig.3 as a horizontal line. Numerical results agree with this analytical prediction within the error range [28].

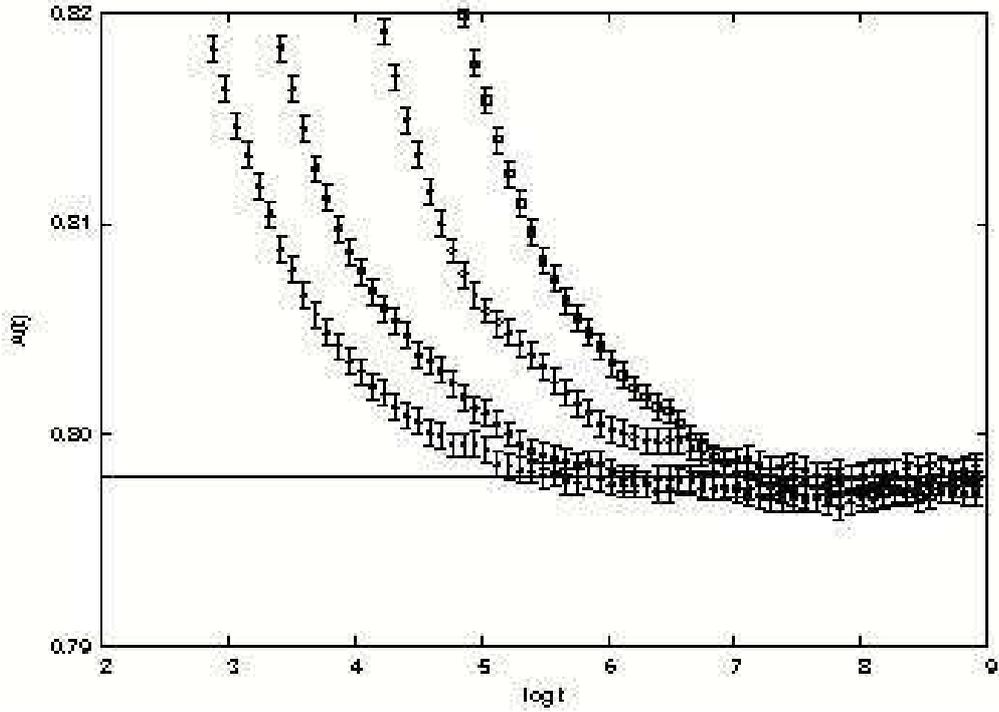

**Fig.3.** The behavior of $A(t) = N(t)/t^\alpha$ as a function of time on a logarithmic scale for four values of $q = 0.5, 0.9, 0.99$ and $0.999$ (from left to right).

## VII. SUMMARY

In uncorrelated subdiffusive CTRW (jump model) the independent waiting times have a probability decaying as a power law with the exponent $\alpha < 1$, thus the mean waiting time diverges, while the second moment of the jump length distribution is finite. As the result, the mean squared displacement grows as $t^\alpha$. In our paper we consider the generalization of CTRW model by including correlations between waiting times. We use the Langevin approach to the CTRW and generalize it by introducing a memory in the Langevin equation for temporal variable and obtain the general expression for the MSD as a function of time, Eq.(31). We consider two generic types of correlations between the time increments, namely short – range exponential correlations and long – range power – law ones. For exponential correlations with typical range $\Delta$ the mean squared displacement grows as $t^{\alpha/(1+\alpha)}$ at short times, $t < \Delta$, whereas at long times, $t > \Delta$, it exhibits the same behavior as in the uncorrelated case, that is grows as $t^{\alpha}$. For the power – law correlations decaying with the exponent $\beta$, $0 < \beta < 1$, the MSD grows as $t^{\alpha/\nu}$, where $\nu = 1 + \alpha(1-\beta)$. In the case of very strong correlations, $\beta = 0$, the diffusion exponent is bounded from below by $\alpha/(1 + \alpha)$, which coincides with the short time exponent in case of exponential correlations. Thus, in the limiting case of normal diffusion, $\alpha = 1$, and very strong correlations we get 1/2 for the diffusion exponent. We note that our approach naturally allows us to include correlations between jump length by the use of non – white noise in the Langevin equation for spatial variable.

**Acknowledgements.** AVC acknowledges the support from European Commission under MC IIF grant No.219966 (LeFrac) as well as by DAAD. IMS acknowledges partial financial support by DFG within SFB 555 Research Collaboration Program.

# APPENDIX A: USEFUL FORMULAS FOR THE MOMENTS OF STABLE DISTRIBUTIONS

**1. General expression for the $q$ – th moment.**

The $q$ – th moment $\langle |x|^q \rangle$ of the random variable $x$ distributed with the PDF $f(x)$ is given by [29]

$$\langle |x|^q \rangle = 2\pi^{-1}\Gamma(1+q)\sin(\pi q/2)\int_0^\infty dk\, k^{-q-1}\left[1 - \text{Re}\,\hat{f}(k)\right], \tag{A.1}$$

where $\hat{f}(k)$ is the corresponding characteristic function. To show this we first note that

$$\int_0^\infty dk\, k^{-q-1}\left(1 - e^{-pk}\right) = p^q q^{-1}\Gamma(1-q) \tag{A.2}$$

(integration by parts). Making an analytical continuation $p = ix$, and using

$$(ix)^q = e^{iq\frac{\pi}{2}\text{sign}(x)}|x|^q, \tag{A.3}$$

we get

$$\int_0^\infty dk\, k^{-q-1}(1 - \cos kx) = |x|^q q^{-1}\Gamma(1-q)\cos\left(\frac{q\pi}{2}\right), \tag{A.4}$$

and

$$|x|^q = 2\pi^{-1}\Gamma(1+q)\sin\left(\frac{q\pi}{2}\right)\int_0^\infty dk\, k^{-q-1}(1-\cos kx) \ . \tag{A.5}$$

Averaging Eq.(A5) over the distribution of $x$ we obtain

$$\langle|x|^q\rangle = \int_{-\infty}^\infty dx\,|x|^q f(x) = 2\pi^{-1}\Gamma(1+q)\sin\left(\frac{\pi q}{2}\right)\int_0^\infty dk\, k^{-q-1}\left[1-\mathrm{Re}\int_{-\infty}^\infty dx\, f(x)e^{ikx}\right] , \tag{A.6}$$

and arrive at Eq.(A1).

## 2. The $q$-th moment of symmetric Lévy stable distribution.

We take the characteristic function of the symmetric stable distribution as

$$\hat{f}(k) = e^{-\sigma^\alpha |k|^\alpha} \ ,\ 0<\alpha<2 , \tag{A.7}$$

where $\sigma$ is a scale parameter.

Then, from Eq.(A1)

$$\langle|x|^q\rangle = 2\pi^{-1}\Gamma(1+q)\sin\left(\frac{\pi q}{2}\right)\sigma^q \int_0^\infty d\xi\, \xi^{-q-1}\left[1-e^{-\xi^\alpha}\right] ,\ q<\alpha. \tag{A.8}$$

Performing integration over $\xi$ we arrive at

$$\left\langle |x|^q \right\rangle = \frac{2}{\pi q}\Gamma(1+q)\Gamma\left(1-\frac{q}{\alpha}\right)\sin\left(\frac{\pi q}{2}\right)\sigma^q \ . \tag{A.9}$$

This result was obtained in Ref.[30].

## 3. Negative moments of one-sided Lévy stable distribution.

Let us calculate

$$\left\langle x^{-q} \right\rangle = \int_0^\infty dx\, x^{-q} L_\alpha(x) \ , \tag{A.10}$$

where $L_\alpha(x)$ is one – sided (totally skewed) alpha – stable Lévy distribution whose Laplace $y \to u$ transform is $\tilde{L}(u) = \exp(-u^\alpha)$.

At first, we use

$$\int_0^\infty ds\, s^{q-1} e^{-xs} = x^{-q}\Gamma(q) \ , \quad q > 0 , \tag{A.11}$$

thus

$$x^{-q} = \frac{1}{\Gamma(q)} \int_0^\infty ds\, s^{q-1} e^{-xs} \ . \tag{A.12}$$

Then,

$$\left\langle x^{-q}\right\rangle = \int_0^\infty dx\, x^{-q} L_\alpha(x) = \frac{1}{\Gamma(q)} \int_0^\infty ds\, s^{q-1} \int_0^\infty dx\, e^{-sx} L_\alpha(x) =$$

$$= \frac{1}{\Gamma(q)} \int_0^\infty ds\, s^{q-1} e^{-s^\alpha} = \frac{1}{\alpha \Gamma(q)} \int_0^\infty d\xi\, \xi^{-1+\frac{q}{\alpha}} e^{-\xi} \quad , \tag{A.13}$$

and finally

$$\left\langle x^{-q}\right\rangle = \frac{\Gamma\!\left(\dfrac{q}{\alpha}\right)}{\alpha \Gamma(q)} \quad . \tag{A.14}$$

In a particular case $q = \alpha$ Eq.(A14) gives

$$\left\langle x^{-\alpha}\right\rangle = \frac{1}{\Gamma(1+\alpha)} \quad . \tag{A.15}$$

# APPENDIX B: CHARACTERISTIC FUNCTION OF THE PROCESS $t(s)$ IN THE CORRELATED CASE, EQ.(25).

Let $\tau(s)$ be a white Lévy noise with the one – sided totally skewed PDF. Its discrete – time approximation consists of a sequence of independent identically distributed random variables $\{\tau_i\}$, $i =$ 1,2,…, N, here N plays the role of a discrete time. The characteristic function $\hat{p}_\tau(k)$ of the random variables $\{\tau_i\}$ is given by

$$\hat{p}_\tau(k) = \exp\left\{-|k|^\alpha \exp\left(-\frac{i\pi\alpha}{2}\operatorname{sgn} k\right)\right\} \quad . \tag{B.1}$$

Then, the characteristic function $\hat{L}_\alpha(k,s)$ of an $\alpha$ - stable totally skewed Lévy motion $t(s) = \int_0^s ds'\,\tau(s')$, see Eq.(6), is obtained with the use of its discrete – time approximation, $t(s) \to \sum_{i=1}^N \tau_i$, as

$$\hat{L}_\alpha(k,s) \to \prod_{i=1}^N \hat{p}_\tau(k) = \exp\left\{-n|k|^\alpha \exp\left(-\frac{i\pi\alpha}{2}\operatorname{sgn} k\right)\right\}$$

$$\to \exp\left\{-s|k|^\alpha \exp\left(-\frac{i\pi\alpha}{2}\operatorname{sgn} k\right)\right\} \quad . \tag{B.2}$$

According to Eq.(23) $t(s) = \int_0^s ds'\,\tau(s')\mu(s')$ (where the first variable of function $\mu$ is omitted), and $\tau(s)$ is a white Lévy noise. The discrete – time approximation to this integral is $t(s) \to \sum_{i=1}^N \tau_i \mu_i$. The characteristic function of each summand is obtained by noting that if some random variable $y$ has a characteristic function $\hat{f}_y(k)$, then the characteristic function of the random variable $z = \mu y$, $\mu = const$, is $\hat{f}_z(k) = \hat{f}_y(\mu k)$. Then, the characteristic function $\hat{p}(k,s)$ of the random process $t(s)$ reads

$$\hat{p}(k,s) \to \prod_{i=1}^n \exp\left(-\mu_i^\alpha |k|^\alpha \exp\left(-\frac{i\pi\alpha}{2}\operatorname{sgn} k\right)\right)$$

$$= \exp\left(-|k|^\alpha \sum_{i=1}^{n} \mu_i^\alpha \exp\left(-\frac{i\pi\alpha}{2}\operatorname{sgn} k\right)\right)$$

$$\to \exp\left[-|k|^\alpha \int_0^s ds' \mu^\alpha(s') \exp\left(-\frac{i\pi\alpha}{2}\operatorname{sgn} k\right)\right] \quad , \qquad (B.3)$$

which result is used in derivation of Eq.(25).

# APPENDIX C: SHORT TIME ASYMPTOTICS OF THE MSD IN CASE OF EXPONENTIAL CORRELATIONS

With the use of Eqs.(16), (27) and (28) we get for the Laplace transform of the MSD,

$$\langle x^2 \rangle(\lambda) = 2\int_0^\infty ds\, s\, \tilde{h}(s,\lambda) = 2\lambda^{\alpha-1}\int_0^\infty ds\, s\, \phi'(s) e^{-\lambda^\alpha \phi(s)} \quad . \qquad (C.1)$$

The integral

$$I(\lambda) = \int_0^\infty ds\, s\, \phi'(s) e^{-\lambda^\alpha \phi(s)} \qquad (C.2)$$

can be evaluated at large $\lambda$ by the use of the Laplace method. Indeed, this integral is of the form

$$I(x) = \int_a^b ds\, e^{-xp(s)} q(s) \quad , \qquad (C.3)$$

with $x = \lambda^\alpha$, $a = 0$, $b = \infty$,

$$p(s) = D_\alpha \phi(s) > p(s = a = 0) = 0, \quad p(s \to 0) \approx \frac{D_\alpha s^{1+\alpha}}{(1+\alpha)\Delta^\alpha}, \tag{C.4}$$

$$q(s) = s\phi'(s), \text{ and } q(s \to 0) \approx \frac{s^{1+\alpha}}{\Delta^\alpha}. \tag{C5}$$

Thus, we see that the conditions of the Theorem 7.1 from Ref.[31] are fulfilled. The integral $I(x)$ can be evaluated as

$$I(x) \approx \frac{Q}{\mu} \Gamma\left(\frac{\lambda}{\mu}\right) \frac{e^{-xp(a)}}{(Px)^{\lambda/\mu}}, \tag{C6}$$

with

$$Q = \frac{1}{\Delta^\alpha}, \quad \lambda = 2+\alpha, \quad P = \frac{1}{(1+\alpha)\Delta^\alpha}, \quad \mu = 1+\alpha. \tag{C7}$$

With the use of Eqs.(C6) and (C7) we get

$$\langle x^2 \rangle(\lambda) = 2\lambda^{\alpha-1} I(\lambda^\alpha) = 2\Delta^{\alpha/(1+\alpha)} (1+\alpha)^{1/(1+\alpha)} \Gamma\left(1 + \frac{1}{1+\alpha}\right) \lambda^{-\frac{1+2\alpha}{1+\alpha}}. \tag{C8}$$

Now, using the Laplace transform pair $L\{t^{k-1}\}(\lambda) = \Gamma(k)/\lambda^k$, we obtain the result for $\langle x^2 \rangle(t)$ which coincides with that obtained by the different method in Section IV, see Eqs.(36) and (37).